# Half-Heusler topological insulators

Binghai Yan[1] and Anne de Visser[2]

[1]Max Planck Institute for Chemical Physics of Solids, Germany;
binghai.yan@cpfs.mpg.de

[2]Van der Waals-Zeeman Institute, Faculty of Science, University of Amsterdam, The Netherlands; a.devisser@uva.nl

**Abstract:** Ternary semiconducting or metallic half-Heusler compounds with an atomic composition 1:1:1 are widely studied for their flexible electronic properties and functionalities. Recently, a new material property of half-Heusler compounds was predicted based on electronic structure calculations: the topological insulator. In topological insulators, the metallic surface states are protected from impurity backscattering due to spin-momentum locking. This opens up new perspectives in engineering multifunctional materials. In this article, we introduce half-Heusler materials from the crystallographic and electronic structure point of view. We present an effective model Hamiltonian from which the topological state can be derived, notably from a non-trivial inverted band structure. We discuss general implications of the inverted band structure with a focus on the detection of the topological surface states in experiments by reviewing several exemplary materials. Special attention is given to superconducting half-Heusler materials, which have attracted ample attention as a platform for non-centrosymmetric and topological superconductivity.

**Taxonomy:** crystallographic structure, electronic material, electronic structure, electrical properties, superconducting

## Introduction

The discovery of a new class of materials, topological insulators (TIs), has ignited intense research activity in the condensed matter physics and materials science communities. TIs have the extraordinary property that the interior of the material is insulating, while the boundaries host exotic metallic surface states as a result of a non-trivial topology in the band structure.[1,2] The existence of TIs was predicted from the electronic band structure of classes of crystalline insulators for generic symmetries, such as time-reversal and charge conjugation symmetries (see Reference 3 for a review) (see the Introductory article in this issue). Solving the Hamiltonian near the surface of the insulator produces the unique feature of TIs: gapless boundary states with extended wave functions that are protected against deformation or perturbations as long as the topology of the bulk band structure that is characterized by a topological index, called $Z_2$ [1,2], is preserved. The topological protection of the metallic surface states is a novel phenomenon in the domain of quantum electronic materials and has generated much excitement in view of its exploitation in fields ranging from spintronics and magnetoelectrics to quantum computation. Topological insulating band order was first predicted and realized in two-dimensional (2D) quantum wells derived from the semimetal HgTe,[4,5] but with the predictions and realization of three-dimensional (3D) topological insulators,[6–8] the resources for engineering new TIs with concurrent functionalities become almost unlimited. While a 2D TI is usually a thin film with metallic topological edge states, a 3D TI is an insulating bulk material that exhibits metallic topological surface states on all outer surfaces.

The (Bi,Sb)$_2$(Te,Se)$_3$ type of 3D TIs[9–12] are currently the most extensively studied TI materials. A crucial ingredient here is the presence of a "heavy" element, e.g. an element in period-4 or above, that exhibits strong spin–orbit coupling (SOC). The strong SOC gives rise to an additional level crossing (band inversion) that drives the system into a topological phase.. In this review, we focus on half-Heusler topological insulators. Half-Heusler compounds form a vast group of cubic ternary crystalline materials with composition *XYZ* that are derived from cubic *X$_2$YZ* compounds named after their discoverer, Fritz Heusler.[13] (Half-)Heusler compounds have attracted ample attention not only as multifunctional materials in the fields of spintronics and thermoelectricity, but also as tunable laboratory tools to study a wide range of intriguing physical phenomena, such as half-metallic magnetism, giant magnetoresistance, and Kondo or heavy-fermion physics (for a review see Reference 14). We now may add TI phenomena to this list.

Recently, *ab initio* electronic structure calculations on a series of "heavy-element" half-Heuslers[15,16] show that these materials are zero-gap semi-metals, but with a band inversion like that observed in HgTe. The real TI state may subsequently be realized, for instance, by applying or engineering strain. The wide range of possible composition variations in the half-Heusler crystal structure is expected to allow for highly tunable and versatile electronic properties, such as spin-structured topological surface states and topological superconductivity. Indeed, some of the topological half-Heusler materials turn superconducting at liquid helium temperatures: LaPtBi,[17] YPtBi,[18] LuPtBi,[19] and ErPdBi.[20] A hot topic in the field of topological superconductors is the realization of Majorana fermion states, which are predicted to exist as protected bound states at the edges of the material.[1,2,21] A Majorana fermion is a particle that is identical to its own antiparticle. Majorana fermions can be bound to a defect at zero energy and then are called the zero modes or bound states. In a 1D superconductor, the Majorana zero mode has non-Abelian statistics, such that adiabatically exchanging particles noncommutatively changes the system from one ground state to another providing a unique platform for topological quantum computation.

The purpose of this article is to review the half-Heusler topological insulators. We start with a brief introduction to half-Heusler materials based on the crystallographic and electronic structure. Next, we formulate the concepts of topological band theory and derive the non-trivial band structure. We apply the theory to the topological semimetal HgTe and subsequently extend the model to 3D half-Heuslers. Then we turn to the topological surface states and their experimental signature. We close by reviewing candidate topological superconductors taken from the half-Heusler series.

**Half-Heuslers: Crystal structure and electronic structure**

Today, the class of Heusler materials comprises more than 1500 different compounds (see Reference 14 and the references therein). Semiconducting behavior is mainly found in the subclass of the half-Heusler $XYZ$ compounds. A half-Heusler material is a crystalline compound consisting of a covalent and an ionic part: the $X$ and $Y$ atoms have a distinct cationic character, whereas $Z$ can be seen as the anionic counterpart. The most electropositive element *X* (usually a main group element, a transition metal, or a rare-earth element) is placed at the beginning of the formula. Generally, the half-Heusler phases crystallize in a non-centrosymmetric structure corresponding to the space group $F\bar{4}3m$. Within the lattice, atoms *Y* and *Z* form a covalent-type ZnS structure, while *X* and *Z* form the ionic NaCl-type substructure, as shown in **Figure 1**. Usually the crystal structure can be viewed as stuffing interstitial sites in a zinc-blende lattice (e.g., GaAs or HgTe) with a third element. For example, the *XYZ* crystal can be understood as stuffing the $(YZ)^{n-}$ zinc-blende sublattice with an $X^{n+}$ ion.

The electronic properties of half-Heusler compounds can be predicted by counting the number of valence electrons. When the total number of valence electrons per formula unit is 8 or 18 (for the closed shell), dubbed as the 8 or 18-electron-rule, half-Heusler *XYZ* compounds can exhibit semiconducting properties that are close to those of classical semiconductors, such as GaAs. For example, LiMgAs with 8 valence electrons is a semiconductor with an energy gap ($E_g$) of 2.3 eV,[22] and ScPtSb with 18 valence electrons is found to be a narrow gap semiconductor with $E_g = 0.7$ eV.[23] The electron count is illustrated in Figure 1c.

One of the most attractive aspects of half-Heusler compounds is the tunability of the insulating gap over a wide energy range, from about 4 eV (e.g., LiMgN) down to zero (e.g., ScPtBi), by choosing *XYZ* combinations with different electronegativity of the constituents and with different lattice constants. Because of the similarity to the TI HgTe in both structural and electronic properties, Heusler materials offer a platform for the design of new TIs by realizing band structures that look similar and are actually topologically equivalent to that of HgTe.

**Topological bulk states, the prototype TI HgTe**

For understanding the topology of the bulk band structure of a half-Heusler, the electronic structure of HgTe may serve as the starting point. The topology of the bulk band structure is characterized by an inverted order between a valence band and a conduction band that have opposite parities.[4] The band order is inverted with respect to that of a normal insulator, such as CdTe. The energy bands nearest to the Fermi level in HgTe and CdTe are located around the Γ-point in the Brillouin zone, as shown in **Figure 2**. Like the common semiconductor GaAs, CdTe has a normal band ordering with the $\Gamma_6$ band as the conduction band minimum (CBM) and $\Gamma_8$ as the valence band maximum (VBM). Bands $\Gamma_6$ with "−" parity and total angular momentum $j = 1/2$ and $\Gamma_8$ with "+" parity and $j = 3/2$ are labeled by the symmetry of the wave functions and constituted by Cd-*s* and Te-*p* orbitals, respectively. Compared to Cd, Hg is a heavier element and undergoes stronger relativistic effects. As a result, the Hg-*s* state ($\Gamma_6$) is much lower in energy than the Te-*p* state ($\Gamma_8$). Therefore, HgTe exhibits an "inverted" band ordering: the *p*-type $\Gamma_8$ band is situated above the *s*-type $\Gamma_6$ band (Figure 2, lower left). This kind of band inversion is essential for the determination of the topologically non-trivial band structure. The $\Gamma_6$–$\Gamma_8$ inversion (*s*–*p* inversion) takes place only around the Γ point, while the normal band ordering ($\Gamma_6$ above $\Gamma_8$) is preserved in the other regions of the Brillouin zone. In addition, the Fermi energy crosses the quadruply degenerate $\Gamma_8$ bands due to the cubic symmetry. This leads to a topological semimetal. In order to realize a real TI, an energy gap can be opened either by quantum confinement in lower dimensions, which induces a 2D TI,[4,5] or by applying strain to break the cubic symmetry, which results in a 3D TI.[24,25]

For the electron structure calculations of HgTe-type compounds, we use the Bernevig-Hughes-Zhang (BHZ) type of Hamiltonian [4], which is based on the ***k·p*** perturbation expansion up to the quadratic term,

$$H = \epsilon(\mathbf{k})I_{4\times 4} + \begin{bmatrix} M(\mathbf{k}) & Ak_+ & & 0 \\ Ak_- & -M(\mathbf{k}) & & \\ & & M(\mathbf{k}) & -Ak_- \\ 0 & & -Ak_+ & -M(\mathbf{k}) \end{bmatrix} \quad (1)$$

Here $\epsilon(\mathbf{k}) = C - D\mathbf{k}^2$, $M(\mathbf{k}) = M_0 + B\mathbf{k}^2$, $k_\pm = k_x \pm ik_y$ and $I_{4\times 4}$ is the 4 × 4 unitary matrix, and $2M_0$ corresponds to the energy gap. The states are expressed in the basis $|\Gamma_6^-, m_j = \frac{1}{2}\rangle$, $|\Gamma_8^+, m_j = \frac{3}{2}\rangle$, $|\Gamma_6^-, m_j = -\frac{1}{2}\rangle$ and $|\Gamma_8^-, m_j = -\frac{3}{2}\rangle$, where "+/−" represents the parity eigenvalues, and $m_j$ is the angular momentum projection. Besides for HgTe, the BHZ model has also been generalized to describe other TIs, such as $Bi_2Se_3$-type TIs,[9] the perovskite oxide TI $BaBiO_3$,[26] and the KHgSb-type weak TIs.[27] As we will see, the BHZ model also applies to Heusler TIs, since Heusler materials exhibit the same $\Gamma_6$–$\Gamma_8$ band inversion as HgTe.

For a given material, the parameters in the effective model can be determined by fitting the energy spectrum of the effective Hamiltonian to that of first-principles calculations. However, one can still obtain insight in the band structure without knowing concrete parameter values. Without loss of generality, we can omit the identity term $\epsilon(\mathbf{k})$ and focus on the second 4 × 4 matrix in Equation 1. The diagonal terms $\pm M(\mathbf{k})$ represent $\Gamma_6$ and $\Gamma_8$ bands with electron and hole-like parabolic dispersions, respectively, assuming B > 0. One can find that the energy gap at the $\Gamma$ point is $2M_0$. The off-diagonal linear terms $Ak_\pm$ correspond to the hybridization between $\Gamma_6$ and $\Gamma_8$ bands that have opposite parities. When $M_0 > 0$, the $\Gamma_6$ bands are always higher in energy than the $\Gamma_8$ bands, which correspond to the case of a trivial insulator such as CdTe. In contrast, when $M_0 < 0$, one finds the $\Gamma_6$ bands are lower in energy than $\Gamma_8$ at the $\Gamma$ point, which corresponds to the case of a TI such as HgTe. For the purpose of studying the topological properties, it is sometimes convenient to project the continuum $\mathbf{k}\cdot\mathbf{p}$ model to a lattice model (i.e., a tight-binding representation), which gives the energy spectrum over the entire Brillouin zone. We can obtain a simplified lattice model by replacing $k_{x,y,z}$ and $k_{x,y,z}^2$ in Equation 1 by $\frac{1}{a}\sin(k_{x,y,z}a)$ and $\frac{2}{a^2}(1 - \cos(k_{x,y,z}a))$, respectively, where $a$ is the lattice constant and usually taken as 1 for simplicity. The corresponding bulk band structures are shown in **Figure 3**.

**The topological invariant $Z_2$**

A commonly used criterion to determine whether a crystalline material is a 3D topological insulator is by its $Z_2$-valued topological invariant.[7] A TI is characterized by 4 $Z_2$ invariants: $\nu_0$ and $\nu_1, \nu_2, \nu_3$. The most important index is $\nu_0$: $\nu_0 = 1$ represents a strong TI, while $\nu_0 = 0$ is a trivial insulator (or a weak TI when at least one of the other $\nu_i$ is 1). The calculation of $\nu_0$ is simplified when the crystal structure possesses inversion symmetry,[28]

$$(-1)^{\nu_0} = \prod_{i=1}^{8} \delta^i \qquad (2)$$

where $\delta^i = \prod_{i=1}^{N} \xi_{2n}(K^i)$ is the product of the parity eigenvalues $\xi_{2n}(K^i)$ of all valence bands (with total number $N$) at the $i$th time-reversal invariant $k$-point $K^i$, and the final product runs over all eight $K^i$ points in the Brillouin zone. Each time-reversal conjugate pair (i.e., the $2n$th and ($2n$–1)th bands, only counts once in the parity product. These eight $k$-points are $K^i$ (0/0.5, 0/0.5, 0/0.5) in units of the reciprocal lattice vectors and satisfy $K^i \equiv -K^i$ and $TH(K^i)T^{-1} = H(-K^i) = H(K^i)$ (i.e., the Hamiltonian remains invariant under the time-reversal symmetry operation $T$ at $K^i$). Even for systems without inversion symmetry, one can apply this criterion by adiabatically changing the material to a centrosymmetric system. For example, we can assign the parity eigenvalues to the HgTe wave functions by adiabatically connecting HgTe to the diamond lattice. For the model in Equation 1, $\nu_0 = 0$ for $M_0 > 0$ because all valence bands are "+" in parity for all eight $K^i$. However, $\nu_0 = 1$ for $M_0 < 0$ because the valence band contributes a "–" sign at the $\Gamma$ point, while the other seven $K^i$ points are "+." Consequently, HgTe is found to be a TI due to the band inversion, as illustrated in Figure 3.

**Half-Heusler topological insulators**

The diversity of Heusler materials allows one to find topological non-trivial band structures similar to HgTe. Based on first-principles calculations, tens of half-Heusler compounds show inverted band structures,[15,16] as illustrated in **Figure 4**. For example, ScPtBi exhibits band inversion similar to HgTe, while ScPtSb displays a normal band ordering similar to CdTe, as shown in Figure 2. Hence we conclude that ScPtBi is a TI, while ScPtSb is a normal insulator. Here the conclusion based on the $\Gamma_6$–$\Gamma_8$ band inversion is consistent with that of directly calculating the $Z_2$ invariant.[29] First-principles calculations using the optimized exchange-correlation potential[30,31] and GW approach[32], a method to calculate the self-energy of many-body systems where G stands for the Green's function and W for the dynamically screened Coulomb potential, have confirmed that a large number of half-Heuslers are TIs with inverted band structures, and at the same time refined some of the topological features at the topological phase transition boundary.

Since the *s-p* band inversion determines the topological nature of HgTe and half-Heuslers, the interesting question arises as to what kind of mechanism produces the inversion. For the (Bi,Sb)$_2$(Se,Te)$_3$ family of TIs, the answer is clear.[9,10] The bulk energy gap usually opens without SOC. By switching on SOC in calculations, the original gap closes, and a new energy gap is re-opened by inverting the conduction and valence bands. For Heusler compounds and HgTe, however, band structure calculations commonly show an *s-p* inverted band structure without SOC.[33] This inversion can be attributed to other relativistic effects: the mass-velocity and Darwin corrections that are commonly considered in state-of-the-art density-functional theory calculations. One can understand these effects in the following simple picture.[34]

Inside an atom of the heavy element, the inner electrons move very fast, with speeds close to the speed of light. Einstein's theory of relativity tells us that this leads to a mass (*m*) increase. The increased mass induces a smaller Bohr radius $a_0$ (e.g., $a_0 = \hbar/me^2$ for a hydrogen-like atom). This leads to a relativistic orbital contraction and energy decrease of all *s* and most of the *p* orbitals, the so-called direct relativistic effect. Moreover, because the contraction of the inner *s* and *p* shells results in strong screening of the nuclear charge, the outer *d* and *f* shells expand and destabilize, which reduces the nuclear screening. As an indirect consequence, the modified nuclear shielding stabilizes the outermost valence *s* shell (e.g., Hg-5*s*) by the enhanced attractive potential. Thus, both direct and indirect relativistic effects, which are included as mass-velocity and Darwin corrections in density-functional theory calculations, reduce the energy of the valence *s* electrons in heavy elements. Thus the *s-p* band inversion in HgTe and half- Heuslers is due to relativistic effects, just like SOC in the (Bi,Sb)$_2$(Se,Te)$_3$-type systems.

**Topological surface states**

One of the hallmarks of topological surface states is the linear energy dispersion (Dirac cone), similar to that in graphene. Moreover, the spin is locked to the momentum, which leads to a helical spin texture inside the Dirac cone. The ideal experimental tool to measure the Dirac cone and spin structure is angle-resolved photoemission spectroscopy (ARPES).[1] In addition, other surface sensitive techniques can be used, such as scanning tunneling microscopy and spectroscopy (STM and STS).[35] Another powerful experimental tool is magnetotransport, i.e. resistance measurements in a magnetic field (see Reference 12). Quantum oscillations in the resistance of the material as a function of the magnetic field (the Shubnikov—de Haas effect) may identify the surface states and their topological nature by inspecting the phase of the oscillations (via the geometrical phase or Berry's phase. However, since magnetotransport probes the interior of the sample (the bulk) *and* the surface, prerequisites are a low bulk conductivity and a high transport mobility of the surface carriers. Another way to investigate the presence of Dirac dispersion of the surface states is by resistance measurements in very high magnetic fields, which is predicted to have a linear field variation.[36]

Although the topological nature of bulk half-Heusler materials has been put on a firm footing by theory, providing conclusive experimental evidence remains a challenging task. Hitherto, ARPES experiments were performed on LnPtBi, where Ln = Lu, Dy, and Gd.[37] According to the calculations, LuPtBi should be a strong TI (see Figure 4). For the measured [111] surface, metallic states were observed. However, the issue of whether these states are topological in nature could not be settled because of their complexity. The complexity is possibly due to the coexistence of many trivial dangling bond states near the Fermi energy. By counting how often the surface states cross the Fermi energy, one may distinguish topological (odd number of crossings) and trivial (even number of crossings) surface states. This simple rule (Reference 1) has been applied to the surface states of, for example, $Bi_{1-x}Sb_x$.[8] However, it requires a finite bulk energy gap, while the energy gap of unstrained half-Heusler materials is zero. Thus, one cannot naively employ the same rule to the half-Heusler surface.

High-field magnetoresistance measurements have been carried out on a number of compounds, including YPtSb, LuPdSb, LaPtBi, YPtBi,[38,39] and YPdBi.[40] These materials all exhibit a non-saturating magnetoresistance with a linear high-field term and a high charge-carrier mobility, which is in line with linear dispersion and topological surface states. In the case of LaPtBi,[41] YPtBi,[18] and YPdBi,[40] Shubnikov-de Haas oscillations have been reported, but the angular variation of the oscillation frequency shows these originate from small 3D Fermi surface volumes (pockets) and not from surface states.

We can obtain some useful insights in half-Heusler TIs from previous experimental and theoretical work on the HgTe surface, considering the close similarities between half-Heuslers and HgTe in both crystal structure and band structure. Under a relative large strain to break the cubic symmetry, the HgTe surface was found to indeed exhibit Dirac-cone-like topological surface states inside the bulk energy gap by calculations[42–45] that are also sensitive to the surface terminations.[43,45,46] Under a small strain and no strain, however, it is revealed by both the *k·p* model Hamiltonian[44] and first-principle[45] calculations that the topological surface states are buried inside the bulk valence bands, with the Dirac point lying below the VBM at the Γ point. Although they strongly hybridize with bulk bands, it is still possible to distinguish these surface states in recent ARPES experiments.[47,48] Regarding the strain-free Heusler TIs measured by ARPES, we speculate that the surface Dirac cone also merges into the bulk valence bands in these compounds. Thus, one may find the topological surface states near the Γ point inside the valence bands (i.e., below the Fermi energy rather than at the Fermi surface). This calls for accurate surface band-structure calculations and high-resolution ARPES experiments to definitely settle the topological nature of half-Heuslers.

**Candidate topological superconductors**

The possibility of realizing a topological superconductor has attracted much attention. Topological superconductors are closely related to TIs due to the direct analogy between topological band theory and superconductivity: the Bogoliubov-de Gennes Hamiltonian for the quasiparticles of a superconductor resembles the Hamiltonian of a band insulator, with the superconducting gap in the role of the gap of the band insulator.[1,2] Topological superconductors are predicted to be unconventional superconductors with mixed even and odd parity Cooper-pair states, which have surface/edge states that are robust against perturbation.[49] Unfortunately, only a few topological superconductors have been proposed to date (e.g., $Cu_xBi_2Se_3$[50] and $In_xSn_{1-x}Te$),[51] but the experimental evidence is not conclusive, and the search remains ongoing. In this light, it is important to examine the topological aspects of superconductivity in the half-Heusler series.

**Table I** lists candidate topological superconductors, which all belong to the platinum and palladium bismuthide series. The electrical resistivity shows, in all cases, weak metallic- or semi-metallic-like behavior with a broad maximum in the temperature range from 50–100 K, as illustrated in **Figure 5** for YPtBi[18,52] and ErPdBi.[20] All of the compounds listed in Table I are low carrier systems, with a carrier concentration of $n \sim 10^{19}$ cm$^{-3}$ at low temperatures. The superconducting transition temperature, $T_c$, ranges from 0.8 to 1.7 K.

A potential way to identify unconventional superconductivity is via the suppression of $T_c$ in a magnetic field (i.e., by upper critical field measurements), $B_{c2}(T)$. In general, a magnetic field will suppress the superconducting state via the interaction with the spins of the Cooper pair (Pauli limit, $B^P$) and its orbital momentum (orbital limit, $B_{c2}^{orb}$).[53] If both contributions are present, $B_{c2}(T \to 0)$ for a standard spin-singlet weak-coupling BCS superconductor is given by $B_{c2}(0) = B_{c2}^{orb}(0)/\sqrt{1+\alpha^2}$, where $\alpha = \sqrt{2}B_{c2}^{orb}(0)/B^P$ is the Maki parameter. This tells us the measured critical field should be lower than $B_{c2}^{orb}$. An extensive study of YPtBi[54] shows its upper critical field is clearly at odds with this simple scenario, as shown in **Figure 6**, where $B_{c2}(T)$ has been traced in reduced coordinates. In fact, the data also exceed the values of model calculations for an unconventional odd-parity Cooper pair state.[55] This calls for more theoretical work to delineate the intricate behavior in a magnetic field of mixed even and odd-parity Cooper pair states in topological superconductors.

We note that the half-Heusler compounds lack inversion symmetry and therefore may also be classified as non-centrosymmetric superconductors.[56] Here the lack of an inversion center in the crystal structure causes an electric field gradient, leading to the Rashba-type SOC effect, i.e. momentum-dependent spin-splitting of bands. This results in a splitting of the Fermi surface, which impedes spin-singlet or spin-triplet Cooper pairing of the conventional type. Instead, new pairing states, notably mixed even and odd parity Cooper pair states, are predicted to make up the superconducting condensate, just as in topological superconductors.

Yet another surprising observation is made for the most recently discovered member in Table I, ErPdBi.[20] Here the rare-earth atom carries a magnetic moment. The magnetic susceptibility follows a local-moment Curie-Weiss law with an effective moment close to the $Er^{3+}$ free-ion value of 9.58 $\mu_B$ and a small paramagnetic Curie temperature $\theta_P$ = –4.6 K. Transport and magnetic measurements (insert Ref.20) show that the system orders antiferromagnetically at $T_N$ = 1.1 K. Nevertheless, ErPdBi becomes superconducting below $T_c$ = 1.2 K, as shown in Figure 5. The coexistence of superconductivity and antiferromagnetic order is unusual. The magnetic and superconducting phase diagram for this compound is shown in **Figure 7**. Since $T_N \approx T_c$, the interaction of superconductivity and magnetic order is expected to give rise to a complex ground state. Electronic structure calculations reveal a topologically non-trivial band inversion.[20] Accordingly, ErPdBi is advocated as a novel, unique platform to study the interplay of topological states, superconductivity, and magnetic order.

**Concluding remarks**

We have addressed a new functionality in the class of half-Heusler compounds: the topological insulating state. Electronic structure calculations provide solid grounds for a strong band inversion in selected materials that give rise to non-trivial, topological electron states. Detection of the topological state is not straightforward since the surface Dirac cone has probably merged with the bulk valence band. This calls for challenging experimental solutions, such as strain-engineering, to induce a TI gap and/or high-resolution ARPES. The great advantage of ternary Heusler compounds is the wide tunability of the electronic structure that can easily be accomplished by varying the constituents. For instance many of the ternary zero-gap semiconductors may also be formed with a magnetic rare-earth element (for instance ErPdBi), which offers new routes to explore the interplay of magnetism and topological phases. Likewise, some of the band-inverted half-Heuslers turn superconducting at low temperatures, which presents a new playground to investigate topological superconductivity.

**Acknowledgments**

B.Y. acknowledges support from the Max-Planck Institute for the Physics of Complex Systems, helpful discussions with C. Felser, and financial support by an ERC Advanced Grant (291472). The work of A.dV. was carried out in the research program on Topological Insulators of FOM (Dutch Foundation for Fundamental Research of Matter).

**Figures and Captions**

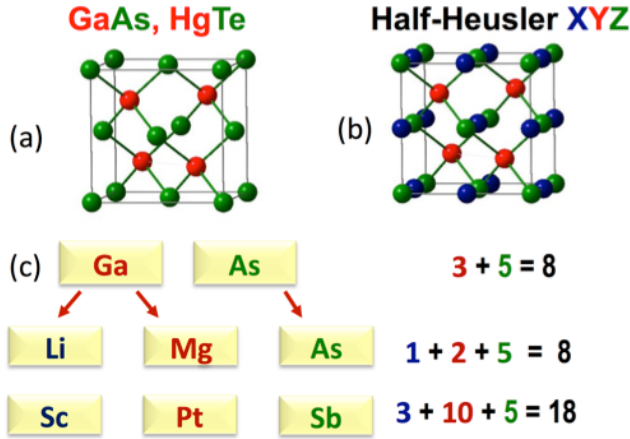

**Figure 1.** Crystal structure of (a) zinc-blende and (b) equiatomic (filled zinc-blende) half-Heusler compounds *XYZ*. Green spheres represent *Z* atoms, red spheres *Y* atoms, blue spheres *X* atoms. (c) Illustration of the 8- or 18-electron rule. Half-Heuslers that fulfill this rule are usually nonmagnetic semiconductors, providing a diverse family of topological insulator (TI) candidates.

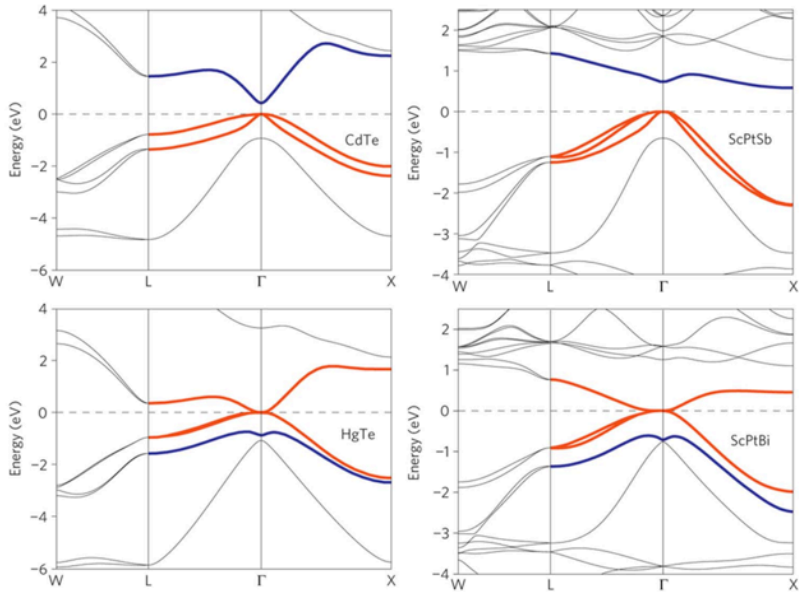

**Figure 2.** Comparison of the electronic band structure of CdTe and HgTe with calculations for the half-Heusler alloys, ScPtSb and ScPtBi. Red color indicates bands with $\Gamma_8$ symmetry, blue with $\Gamma_6$ symmetry. The comparison reveals close similarity between the binary systems and their ternary equivalents: both CdTe and ScPtSb are trivial insulators, with $\Gamma_6$ situated above $\Gamma_8$, which sits at the Fermi energy (set to zero). Both HgTe and ScPtBi are topological insulators with inverted band order; the band with $\Gamma_6$ symmetry is situated below $\Gamma_8$. Figure adapted from Reference 15.

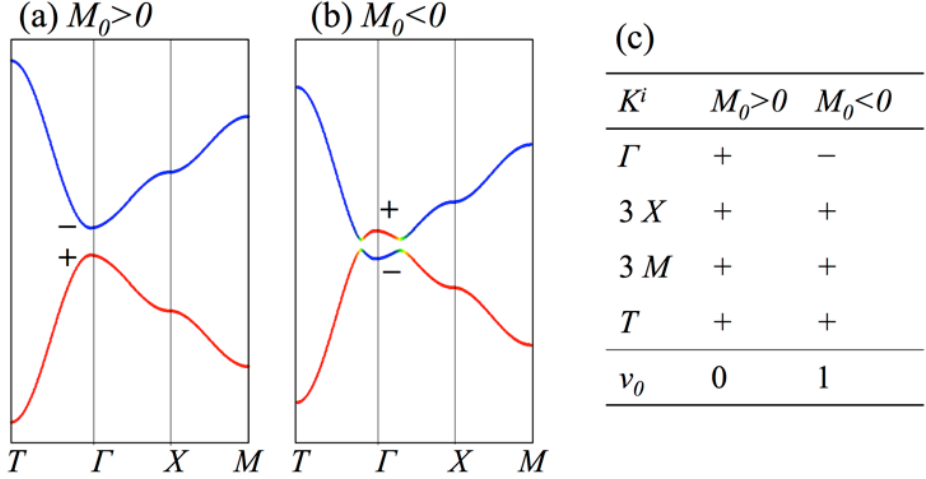

**Figure 3.** Illustration of band structures of (a) a trivial insulator ($M_0 > 0$) and (b) a topological insulator ($M_0 < 0$). The band dispersion is obtained from a lattice model based on the effective Hamiltonian in Eq.1. Blue and red lines represent $\Gamma_6$ ("−" parity) and $\Gamma_8$ ("+" parity) bands, respectively. A band inversion exists clearly for $M_0 < 0$. (c) The parity eigenvalues at the $\Gamma$ (0 0 0) point, the 3 $X$ (0.5 0 0) points, the 3 $M$ (0.5 0.5 0) points and the $T$ (0.5 0.5 0.5) point. The last line lists the topological $Z_2$ invariant $\nu_0$ that can be estimated from Equation 2. The non-trivial $Z_2$ invariant is attributed to the inverted band structure in (b).

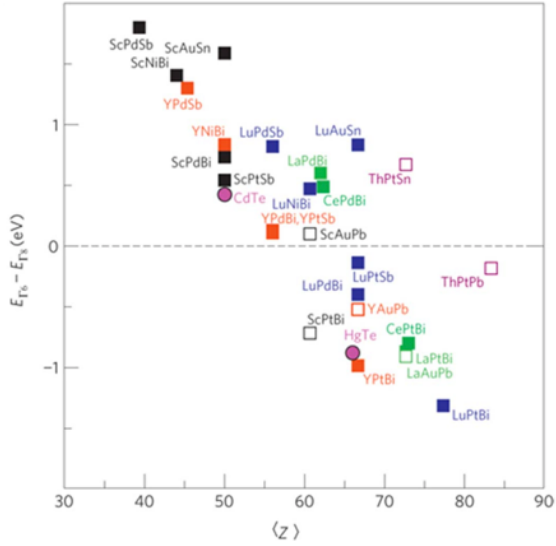

**Figure 4.** The band inversion strength calculated by the difference $\Gamma_6$–$\Gamma_8$ for various half-Heuslers as a function of the average spin–orbit coupling (SOC) strength represented by the average nuclear charge $<Z> = \frac{1}{N}\sum_{i=1}^{N} Z(x)$, where $Z(x)$ is the nuclear charge, and $N$ is the number of elements in the compound. HgTe and CdTe binaries are shown for comparison. Open squares mark the systems not synthesized in experiments As elements become heavier with increasing SOC, the band inversion strength increases. Figure adapted from Reference 15.

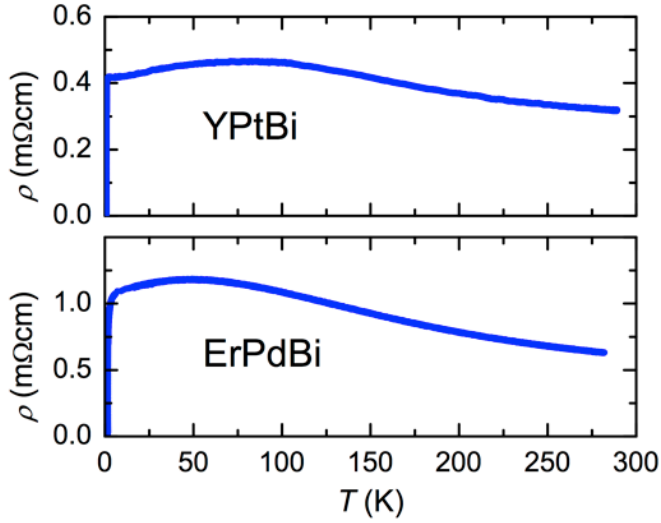

**Figure 5.** Semi-metallic-like electrical resistivity as a function of temperature of YPtBi (upper frame)[52] and ErPdBi (lower frame).[20] Superconductivity is observed near 1 K.

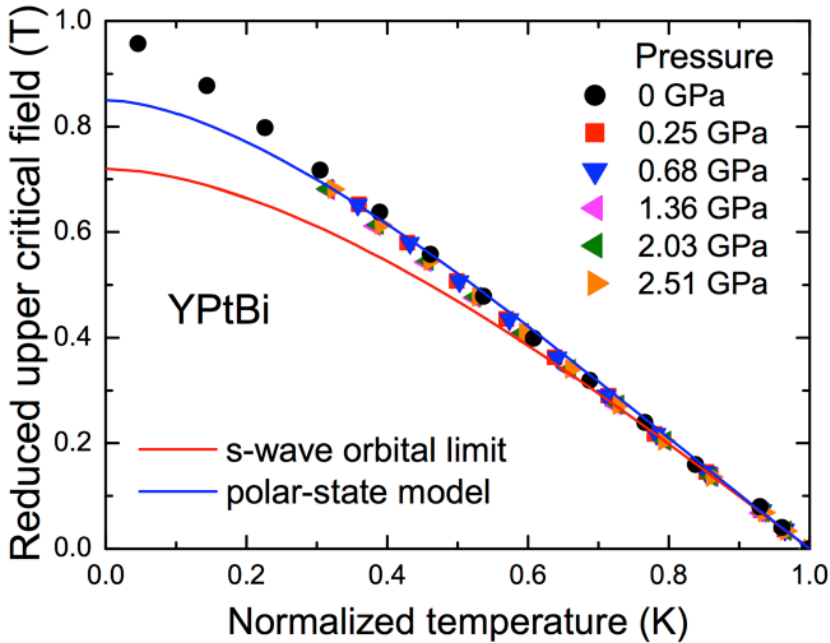

**Figure 6.** The upper critical field of YPtBi measured at ambient pressure and under pressure up to 2.51 GPa as indicated by the different symbols. In reduced coordinates $(B_{c2}(T)/T_c)/(dB_{c2}/dT)|_{T_c}$ versus $T/T_c$, all data collapse onto a universal curve, which exceeds the values for a standard weak-coupling orbital-limited s-

wave superconductor (red line). The blue solid line is a model calculation for a $p$-wave superconductor (see text). See Reference 54 for details.

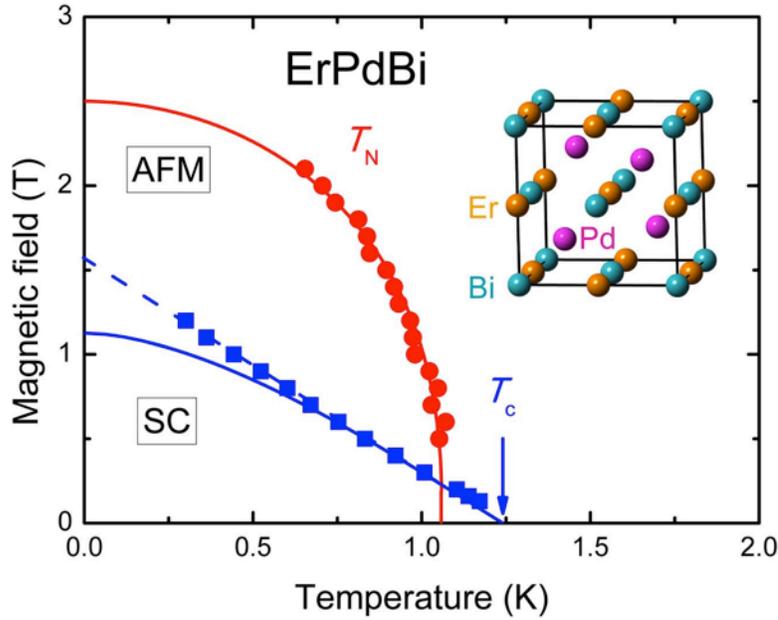

**Figure 7.** Superconducting ($T_c$-SC) and antiferromagnetic ($T_N$-AFM) phase diagram of ErPdBi. The upper critical field (blue squares) shows an unusual linear behavior with values that exceed model calculations for a standard weak-coupling orbital-limited $s$-wave superconductor (blue line). The inset shows the half-Heusler crystal structure with Er, Pd, and Bi atoms. Figure adapted from Reference 20.

| Material | $T_c$ (K) | $B_{c2}$ (T) | Reference |
|---|---|---|---|
| LaPtBi | 0.9 | 1.2 | [17] |
| YPtBi | 0.8 | 1.3 | [18] |
| LuPtBi | 1.0 | 1.6 | [19] |
| ErPdBi | 1.2 | 1.6 | [20] |
| CePdBi* | 1.3 | 1.4 | [57] |

* superconducting volume fraction of only 8%.

**Table I.** Superconducting transition temperature $T_c$ and upper critical field $B_{c2}$ of half-Heusler platinum and palladium bismuthides.